\documentclass[aip,cha,amsmath,amssymb,reprint,author-numerical]{revtex4-1}
\usepackage{graphicx,bm,amsmath,amssymb,natbib,url,epsfig}
\usepackage{dcolumn}
\usepackage{hyperref}
\usepackage{amsmath,amssymb}
\usepackage{xcolor,colortbl}
\usepackage{color}
\usepackage{color}
\definecolor{darkgreen}{rgb}{0.1,.6,.1}
\definecolor{greenblue}{rgb}{0.0,.1,.4}

\usepackage[normalem]{ulem}

\begin{document}

\title{Networks of coupled oscillators: from phase to amplitude chimeras} 

\author{Tanmoy Banerjee}
\email{tbanerjee@phys.buruniv.ac.in}
\affiliation{Chaos and Complex Systems Research Laboratory, Department of Physics, University of Burdwan, Burdwan 713 104, West Bengal, India.}
\author{Debabrata Biswas}
\thanks{Equal contribution}
\affiliation{Department of Physics, Rampurhat College, Birbhum 731 224, West Bengal, India.}
\author{Debarati Ghosh}
\thanks{Equal contribution}
\affiliation{Chaos and Complex Systems Research Laboratory, Department of Physics, University of Burdwan, Burdwan 713 104, West Bengal, India.}
\author{Eckehard Sch{\"{o}}ll}
\email{schoell@physik.tu-berlin.de}
\affiliation{Institut f{\"u}r Theoretische Physik, Technische Universit\"at Berlin, Hardenbergstra\ss{}e 36, 10623 Berlin, Germany.}%
\author{Anna Zakharova}
\affiliation{Institut f{\"u}r Theoretische Physik, Technische Universit\"at Berlin, Hardenbergstra\ss{}e 36, 10623 Berlin, Germany.}
\thanks{Equal contribution}

\received{}
\date{\today}

\begin{abstract}
We show that amplitude-mediated phase chimeras and amplitude chimeras can occur in the same network of nonlocally coupled identical oscillators. These are two different partial synchronization patterns, where spatially coherent domains coexist with incoherent domains and coherence/incoherence refer to both amplitude and phase or only the amplitude of the oscillators, respectively. By changing the coupling strength  the two types of chimera patterns can be induced. We find numerically that the amplitude chimeras are not short-living transients but can have a long lifetime. Also, we observe variants of the amplitude chimeras with quasiperiodic temporal oscillations. We provide a qualitative explanation of the observed phenomena in the light of symmetry breaking bifurcation scenarios. We believe that this study will shed light on the connection between two disparate chimera states having different symmetry-breaking properties. 
\end{abstract}

\pacs{05.45.Xt}
\keywords{Chimera, amplitude-mediated phase chimera, amplitude chimera, coupled oscillators, partial synchronization}
\maketitle 
\begin{quotation}
Chimera states are emergent dynamical patterns in networks of coupled oscillators where coherent and incoherent domains coexist due to spontaneous symmetry-breaking. In oscillators that exhibit both phase and amplitude dynamics, two types of distinct chimera patterns exist, namely, amplitude-mediated phase chimeras (AMC) and amplitude chimeras (AC). In the AMC state coherent and incoherent regions are distinguished by different mean phase velocities: all coherent oscillators have the same phase velocity, however, the incoherent oscillators have disparate phase velocities. In contrast to AMC, in the AC state all the oscillators have the same phase velocity, however, the oscillators in the incoherent domain show periodic oscillations with randomly shifted center of mass. Surprisingly, in all the previous studies on chimeras a given network of continuous-time dynamical systems seems to show either AMC or AC: they never occur in the same network. In this paper, for the first time, we identify a network of coupled oscillators where both AMC and AC are observed in the same system, and we also provide a qualitative explanation of the observation based on symmetry-breaking bifurcations. 
\end{quotation}
\section{Introduction}
\label{sec:intro}

The chimera state is a counterintuitive spatiotemporal pattern in oscillator networks that has been in the center of active research over the past decade \cite{chireview,schoell_rev}. This state is generated by the spontaneous breaking of symmetry in the population of coupled {\it identical} oscillators. As a result the network spontaneously splits into at least two incongruous domains, in one domain the neighboring oscillators are synchronized, whereas in another domain the oscillators are desynchronized. After its discovery in phase oscillators by Kuramoto and Battogtokh \cite{KuBa02}, many theoretical studies \cite{st1,st2,chireview,schoell_rev} established the existence of this state. A series of experimental observation of chimera states established that this state is robust in natural and man-made systems. The first experimental observation of chimeras was reported in optical systems \cite{HaMuRo12} and chemical oscillators \cite{TiNkSh12}. Later, chimeras have been observed experimentally in mechanical systems \cite{MaThFo13,KAP14}, electronic~\cite{LAR13,GAM14}, optoelectronic delayed-feedback \cite{LAR15,raj_comp} and electrochemical~\cite{WIC13,WIC14,SCH14a} oscillator systems and Boolean networks~\cite{ROS14a}. Control methods to stabilize chimera have recently been proposed \cite{SIE14c,BIC15,tweezer,delay_ac}. 
Recent studies, both analytical and experimental, explored the occurrence of chimeras in smaller networks \cite{boe15,raj_minimal,smallest,small1,small2,small3}. The notion of chimeras has recently been extended to noise-induced chimera states~\cite{cr_chimera}, and chimera patterns in two- and three-dimensional regular arrays have been explored \cite{LAI09,MAR10,OME12a,XIE15a,MAI15,2dhovel,hiz3d}. 
Chimera patterns have been found in diverse models in nature, such as ecology \cite{csod,lr16}, SQUID metamaterials \cite{expt_meta,HIZ16a}, neuronal systems \cite{gradpre} and quantum systems \cite{schoell_qm}. Recently, chimera states have been identified in continuous media with local coupling \cite{motter_prl}, which opens up the connection of chimeras with fluid dynamics.

After their discovery in phase oscillators \cite{KuBa02} several other types of chimera states have been discovered in systems with coupled phase and amplitude dynamics, but all those chimera patterns are variants of two general chimera states, namely {\it amplitude-mediated phase chimeras} \cite{amc_sethia} and pure {\it amplitude chimeras} \cite{scholl_CD}. In amplitude-mediated phase chimeras (AMC)  incoherent fluctuations occur in both the phase and the amplitude in the incoherent domain; also, in the incoherent domain the temporal evolution of the oscillators is chaotic. On the other hand, amplitude chimeras (AC) were discovered by Zakharova {\em et al.}~\cite{scholl_CD,ZAK15b,TUM17} where all the oscillators have the same phase velocity but they have uncorrelated amplitude fluctuations in the incoherent domain; Also, unlike AMC (or classical phase chimera), the dynamics of all the oscillators in the AC state is periodic. 

Surprisingly, in all the previous studies on chimeras, AMC and AC have not been observed in the same continuous-time network of coupled identical oscillators: a given network seems to show either AMC or AC. For example AMC have been observed in complex Ginzburg-Landau oscillators under nonlocal coupling \cite{amc_sethia}, van der Pol oscillators \cite{vdp1}, FitzHugh-Nagumo models \cite{Omelchenko13,fhn2} and oscillators showing excitability of type-I \cite{hovel_snic} under nonlocal coupling, but no AC patterns appear in those systems. On the other hand, AC appear in nonlocally coupled Stuart-Landau oscillators \cite{scholl_CD,tanCD,ac_noise,delay_ac} and ecological oscillators \cite{csod,lr16}, but AMC have not been observed in those networks. Previously, the possibility of observing two types of chimera states, amplitude and phase chimeras, has been reported for coupled chaotic maps \cite{BOG16, BOG16a}, while for continuous-time chaotic systems only amplitude chimeras have been detected. It has been shown that amplitude chimeras and phase chimeras can switch in time for a network of nonlocally coupled logistic maps~\cite{VAD16} and Henon maps~\cite{SEM17} operating in the chaotic regime.

In this paper we ask the following question: Can both kind of chimeras (i.e., AMC and AC) be observed in the same system? This is a fundamental question in the study of symmetry-breaking in coupled oscillators because of the following facts: {\bf (i)} Unlike AMC, AC has a connection with the oscillation death state, which is a symmetry-breaking state in a network of coupled oscillators where oscillators split into different branches of inhomogeneous steady states \cite{kosprep}. This connection discovered by Zakharova et al. \cite{scholl_CD} is mediated by the ``chimera death" state in which the population of oscillators splits into distinct coexisting domains of spatially coherent oscillation death and spatially incoherent oscillation death (i.e., where the sequence of populated branches of neighboring nodes is completely random in the inhomogeneous steady state) \cite{scholl_CD}. The above distinction has a broad significance in the context of self-organized states in coupled oscillators out of equilibrium. According to the notion introduced by Prigogine \cite{prigogine,goldprigo} there exists four fundamental types of ``dissipative structures" in physical and biological systems, namely, multistability, temporal dissipative structures (in the form of sustained oscillations), spatial dissipative structures (known as Turing patterns) and spatiotemporal structures (in the form of propagating waves). Out of these four dissipative structures, AMC belongs to the spatiotemporal structure and it has no connection with the spatial dissipative structure (or Turing-type bifurcation). 
On the other hand although AC belongs to the spatiotemporal structure, it has a connection to the spatial dissipative structure, namely ``chimera death". Therefore, AC has relevance where inhomogeneity arises out of homogeneity, which is believed to be the underlying mechanism for morphogenesis and cellular differentiation \cite{turing,cell}.
However, the AMC state may account for the observation of partial synchrony in neural activity, like unihemispheric sleep of dolphins and certain migratory birds \cite{RAT00,st2,MaWaLi10,RAT16}, ventricular fibrillation \cite{DaPeSa92}, and power grid networks \cite{grid1}. {\bf (ii)}~In the context of symmetry these two chimeras are distinct. The underlying type of symmetry-breaking in the case of AMC has recently been explored for four globally coupled oscillators (and also verified for optoelectronic oscillators) by Kemeth et al. \cite{kemethprl}. They have identified that AMC arises due to the emergence of the reduced symmetry state ${\bf S_2}^i \times {\bf S_2}^a$, where ${\bf S_2}$ denotes the permutation symmetry ($i$ and $a$ denote instantaneous and average, respectively, see Ref.~\onlinecite{kemethprl} for details). On the other hand it is well known that AC arises due to the breaking of {\it continuous rotational symmetry} \cite{scholl_CD}. Therefore, from very fundamental point of view AMC and AC have different origin, and thus their appearance in the same system is quite significant.

In this paper we discover that AMC and AC can indeed both occur in a network of nonlocally coupled Rayleigh oscillators. This model was proposed by Lord Rayleigh in 1883 to model the appearance of sustained vibrations in acoustics, e.g., in a clarinet \cite{ral}. Later it has been found to be relevant for modeling human limb movement and was used widely in robotics to simulate locomotion \cite{haken}. Remarkably, in our network, we not only observe the simultaneous occurrence of AMC and AC, but a direct transition from AMC to AC is observed with increasing coupling strength for small coupling range. We further numerically assert that, contrary to the Stuart-Landau model, in the Rayleigh model AC is not a transient state, but it is a stable spatiotemporal pattern. Also, we observe an interesting variant of the AC state with quasiperiodic or chaotic temporal oscillations. These findings bridge two apparently disconnected chimera patterns, namely AMC and AC, and establish AC as a stable chimera pattern.

\section{Network of Rayleigh oscillators}
We consider a ring network of $N$ identical Rayleigh oscillators \cite{ral} coupled through a nonlocal matrix coupling. The mathematical model of the network reads
\begin{subequations}
\label{ral}
\begin{align}
\dot{x}_i &=\omega y_i+\frac{\varepsilon}{2P}\sum_{j=i-P}^{i+P}a_{11}(x_j-x_i)+a_{12}(y_j-y_i)\\
\dot{y}_i &=-\omega x_i+\delta (1-{y_i}^2)y_i\nonumber\\
&+\frac{\varepsilon}{2P}\sum_{j=i-P}^{i+P}a_{21}(x_j-x_i)+a_{22}(y_j-y_i)
\end{align}
\end{subequations}
where $x_i, y_i \in \mathbb{R}$,  $i=1,\dots,N$ and all indices are taken modulo $N$, $\omega$ is the linear angular frequency, and $\delta>0$ governs the nonlinear friction. The coupling strength is denoted by $\varepsilon > 0$, and $P \in \mathbb{N}$ represents the number of coupled neighbors to each side. The rotational coupling matrix is defined as $A=\left( \begin{array}{cc} a_{11} & a_{12} \\ a_{21} & a_{22} \end{array} \right)=\left( \begin{array}{cc} \mbox{cos}\phi & \mbox{sin}\phi \\ \mbox{-sin}\phi & \mbox{cos}\phi \end{array} \right)$ with the coupling phase $\phi$. For $\phi=\pi/2$, i.e., $a_{11}=a_{22}=0$, $a_{12}=-a_{21}=1$ the nodes are connected by a pure conjugate coupling, and for $\phi=0$, i.e., $a_{11}=a_{22}=1$, $a_{12}=a_{21}=0$, the coupling is diagonal through similar variables. This type of coupling with a coupling phase is relevant in neuronal and mechanical systems \cite{matrix_mechanics} and was considered earlier in Refs.~\onlinecite{Omelchenko13,hovel_snic} to observe chimeras.

Following the argument in Ref.~\onlinecite{Omelchenko13}, i.e., a phase reduction of Eq.(\ref{ral}) for small coupling strength, and comparison with the  phase lag parameter of coupled Kuramoto phase oscillators, we choose the coupling phase in the rest of the paper as $\phi=\pi/2-0.1$, which is favorable for chimeras and was used earlier in Refs.~\onlinecite{phaselag,Omelchenko13,hovel_snic}. 

\begin{figure}
\centering
\includegraphics[width=0.47\textwidth]{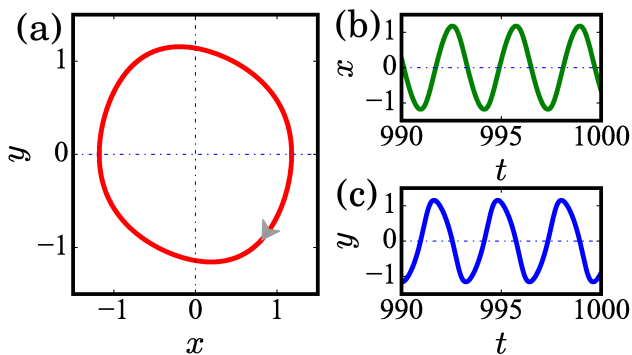}
\caption{\label{fig_ral} (Color online) A single Rayleigh oscillator given by Eq.\ref{ral} with $\varepsilon=0$. (a) Phase portrait of the limit cycle attractor, and
 (b, c) time-series of $x$ and $y$. Parameters are $\omega=2$ and $\delta=1$.}
\end{figure}

\section{Results and analysis}
A single Rayleigh oscillator (Eq.~(\ref{ral}) with $\varepsilon=0$) exhibits a limit cycle oscillation. The frequency and amplitude are determined by $\omega$ and $\delta$ (see Ref.~\onlinecite{haken}). The limit cycle is illustrated in Fig.~\ref{fig_ral} for $\omega=2$ and $\delta=1$. 

\subsection{Spatiotemporal dynamics: Chimera patterns and their characterization}

\begin{figure}
\centering
\includegraphics[width=0.475\textwidth]{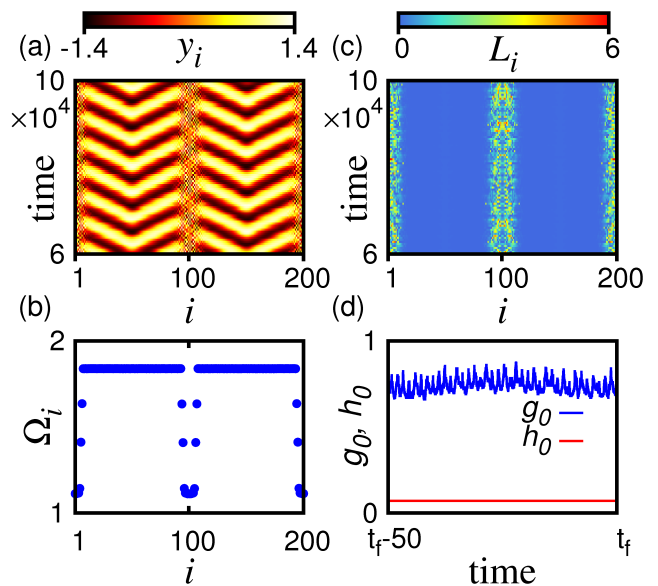}
\caption{\label{fig_amc} (Color online) Amplitude-mediated phase chimera (AMC) for $P=5$ and $\varepsilon=0.8$. (a) Space-time diagram of $y_i$. (b) Mean phase velocity profile $\Omega_i$. (c) Space-time diagram of local curvature $L_i$. (d) Measures of spatial correlation ($g_0$) and temporal correlation ($h_0$) (see text). For clarity only the last 50 time steps are shown ($t_f=10^5$). Other parameters are $\omega=2$ and $\delta=1$, $\phi=\pi/2-0.1$.}
\end{figure}

To demonstrate the observed results clearly, we choose a coupling range $P=5$ and consider two exemplary values of coupling strengths: $\varepsilon=0.8$ for which we observe an AMC state, and $\varepsilon=2$ for which we observe an AC state. Figure \ref{fig_amc}(a) illustrates the space-time pattern of the AMC state for $P=5$ and $\varepsilon=0.8$. One observes two incoherent domains separated by two coherent regions. To ensure that the observed spatiotemporal pattern is indeed an AMC state we use the following characteristic measures: (i) the mean phase velocity profile ($\Omega_i$), (ii) the measure of the local curvature ($L_i$) (iii) the measure of correlation in space ($g_0$) and (iv) the measure of correlation in time ($h_0$); the latter three measures were recently introduced by \citet{kk} as quantitative measures of diverse chimera patterns. In the next few paragraphs we will briefly define and review the properties of these quantities. We define the phase of the $i$-th oscillator as $\psi_i(t)= \mbox{arctan} \left(\frac{y_i(t)}{x_i(t)}\right)$. The {\em mean phase velocity profile} of each oscillator is a good indicator of an AMC state \cite{Omelchenko13} given by
\begin{equation}\label{omi}
\Omega_i=\frac{2\pi M_i}{\Delta T},
\end{equation}
where $M_i$ denotes the number of periods of the $i$-th oscillator in the time interval $\Delta T$. Typically, for an AMC state $\Omega_i$ is flat in the coherent zones and arc-shaped in the incoherent zones. Figure~\ref{fig_amc}(b) shows that in the incoherent domain the mean phase velocity of the oscillators is less than that in the coherent domain, with an arc-shaped profile indicating the occurrence of AMC. 

According to Ref.~\onlinecite{kk}, to find the local curvature at each node $i$ at time $t$ we apply the discrete Laplacian operator $\hat{L}$ on each snapshot $\{\psi_i\}$ that is given by
\begin{equation}
\label{li}
\hat{L}\psi_i(t) = \psi_{(i-1)}(t)-2\psi_i(t)+\psi_{(i+1)}(t).
\end{equation}
If the $i$-th node populates the synchronous cluster, Eq.~\eqref{li} yields $|\hat{L}\psi_i(t)|=0$, but in case of incoherent cluster $|\hat{L}\psi_i(t)|$ is finite. In the incoherent cluster, depending on the phase difference of the neighboring oscillators, $|\hat{L}\psi_i(t)|$ fluctuates between $0<|\hat{L}\psi_i(t)|\leq L_{max}$,  where the maximum local curvature $L_{max}$ is the curvature of nodes having two nearest neighbors with maximum phase shift. Figure~\ref{fig_amc} (c) shows the spatiotemporal variation of $L_i$ corresponding to Fig.~\ref{fig_amc}~(a): it can be seen that in the incoherent domain $L_i$ fluctuates randomly with values $L_i\in(0,6]$, however, in the coherent domains it attains a zero value. At each time step $g(|\hat{L}|=0)$ measures the relative size of the spatially coherent regions, where $g$ represents the normalized probability density function of $|\hat{L}|$. In a fully synchronized system $g(|\hat{L}|=0)=1$, but in case of a completely incoherent system $g(|\hat{L}|=0)=0$ \cite{kk}. Thus, any intermediate value of $g(|\hat{L}|=0)=g_m$, $0<g_m<1$ indicates coexistence of synchronous and asynchronous oscillations. Since spatial coherence and incoherence can only be defined within a certain numerical inaccuracy, we consider a threshold value $\delta_{th}=0.01L_{max}$~\cite{kk} to characterize the coherence or incoherence. Therefore, the spatial correlation measure with the threshold value $\delta_{th}$ is defined as
\begin{equation}
g_0(t) \equiv \sum_{|\hat{L}\psi_i(t)|=0}^{\delta_{th}} g(|\hat{L}\psi_i(t)|).
\end{equation}
To calculate the temporal correlation we consider the pairwise correlation coefficients \cite{kk} defined as 
\begin{equation}
\rho_{ij} \equiv \frac{\left\langle(\psi_i-\left\langle \psi_i \right\rangle)(\psi_j-\left\langle \psi_j \right\rangle)\right\rangle
}{{(\langle\psi_i^2\rangle-\left\langle \psi_i \right\rangle^2)}^{1/2}{(\langle\psi_j^2\rangle-\left\langle \psi_j \right\rangle^2)}^{1/2}},
\end{equation}
here $i\neq j$, $\langle \cdot\rangle$ denotes the temporal mean. With the normalized distribution function $h(|\rho|)$ one can characterize a static $(h(|\rho_{ij}|\approx 1)>0)$ and traveling (or non-static) $(h(|\rho_{ij}|\approx 1)=0)$ spatiotemporal state. The percentage of time-correlated oscillators is defined as,
\begin{equation}
h_0 \equiv \left( \sum_{|\rho|=\gamma}^{1} h(|\rho|)\right)^{1/2}.
\end{equation} 
We consider two oscillators as correlated if $|\rho_{ij}|>0.99=\gamma$. Figure~\ref{fig_amc}~(d) gives the variation of $g_0$ and $h_0$ for the AMC state of Fig.~\ref{fig_amc}~(a): $g_0<1$ ensures the occurrence of chimeras in the network and $h_0>0$ indicates that the resulting chimera is static in nature.
\begin{figure}
\centering
\includegraphics[width=0.48\textwidth]{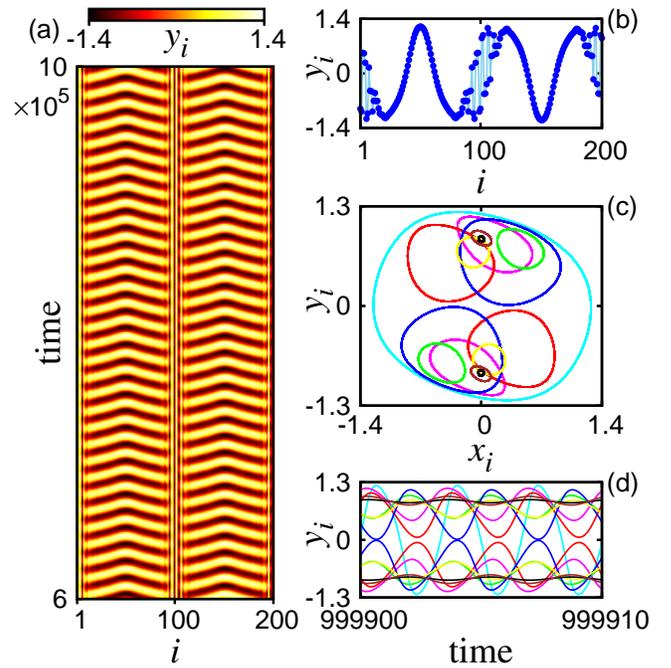}
\caption{\label{fig_ac} (Color online) Amplitude chimera (AC) for $P=5$ and $\varepsilon=2$. (a) Space-time diagram of $y_i$.  (b) Snapshot of $y_i$ at $t=5\times10^5$. (c) Phase portrait in the ($x_i,y_i$) plane of a few selected oscillators: largest cycle (in cyan color) is for an oscillator in the coherent domain, the others are from the incoherent domain. (d) Corresponding time series of $y_i$. Other parameters: $\omega=2$, $\delta=1$, $\phi=\pi/2-0.1$.}
\end{figure}

\begin{figure}
\centering
\includegraphics[width=0.48\textwidth]{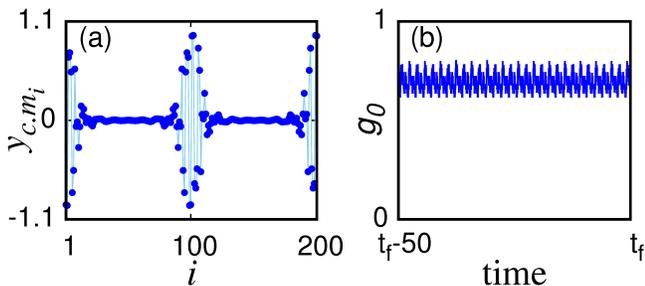}
\caption{\label{cmgh_ac} (Color online) (a) Center of mass ($y_{{c.m}_i}$) of each oscillator corresponding to Fig.~\ref{fig_ac}~(a). Note that in the incoherent domains it shows a random sequence of shifts into the upper and lower halfplane, respectively. (b) Temporal evolution of the spatial correlation measure $g_0$. For clarity only the last 50 time steps are shown ($t_f=10^5$); $g_0<1$ indicates a stable amplitude chimera. Other parameters as in Fig.~\ref{fig_ac}.}
\end{figure}

Next, we demonstrate the occurrence of AC in the network. Figure \ref{fig_ac} (a) shows the spatiotemporal pattern of AC for $P=5$ and $\varepsilon=2$ and Fig.~\ref{fig_ac} (b) shows the corresponding snapshot of $y_i$. The main characteristic feature of an AC is that oscillators exhibit limit cycles with shifted center of mass of the oscillation. This is shown in Fig.~\ref{fig_ac}~(c) using an ($x_i,y_i$) phase portrait for a few representative oscillators selected from the incoherent and the coherent regions, respectively. Figure~\ref{fig_ac} (d) gives the corresponding time series of $y_i$. From the figures it is clear that all the oscillators perform limit cycle oscillations with the same frequency but different amplitude. As a quantitative measure of the AC state we compute the {\em center of mass} of each oscillator \cite{scholl_CD} defined by 
\begin{equation}\label{cm}
y_{{c.m}_i}=\frac{1}{T}\int_0^T y_i dt,
\end{equation}
where $y_i$ represents the state of the $i$-th oscillator and $T$ is a sufficiently large time. The quantity $y_{{c.m}_i}$ gives a measure of the shift of the limit cycle from the origin. Figure~\ref{cmgh_ac}~(a) shows $y_{{c.m}_i}$ of each oscillator, corresponding to Fig.~\ref{fig_ac} (b): we observe that in the incoherent region the center of mass of the oscillators exhibit a random sequence, however, in the coherent region all oscillators have zero center of mass.

A significant observation is that, unlike in previous cases, the amplitude chimera is not a short-living spatiotemporal pattern, rather it has a long lifetime. We check the result for simulation times of $10^7$ and find that the AC pattern does not vanish. We assert that this long lifetime is not a numerical artifact: the same long-living pattern of AC is observed using another integrator that uses the fifth-order Dormand-Prince method of adaptive step size taking absolute tolerance of $10^{-9}$  and relative tolerance $10^{-8}$. This long lifetime is supported by the characteristic measure $g_0$ shown in Fig.~\ref{cmgh_ac}(b), which does not reveal any jump of $g_0$ to a value $1$, rather it fluctuates around $0.7$ for the total time span of our simulation (For clarity only the last 50 time steps are shown ($t_f=10^5$)). To test whether this long-living AC results from the particular initial condition we have used or whether it is a general result of this network, we verify this result for completely random initial conditions (see Appendix A) and find that the AC emerging in this network is indeed a long-living spatiotemporal pattern.

For higher values of the coupling range ($P$) the direct transition from AMC to AC does not occur anymore, instead a multistable state of synchronized oscillations (SYNC) and coherent traveling waves (TW) appears between the AMC and the AC state in parameter space. Further, we observed that the AMC state for the higher $P$ is an {\it imperfect} AMC, i.e., the incoherent domain shows random lateral motion in its spatiotemporal evolution \cite{impchimera,impchimera2,tweezer}. 

\begin{figure}
\centering
\includegraphics[width=0.49\textwidth]{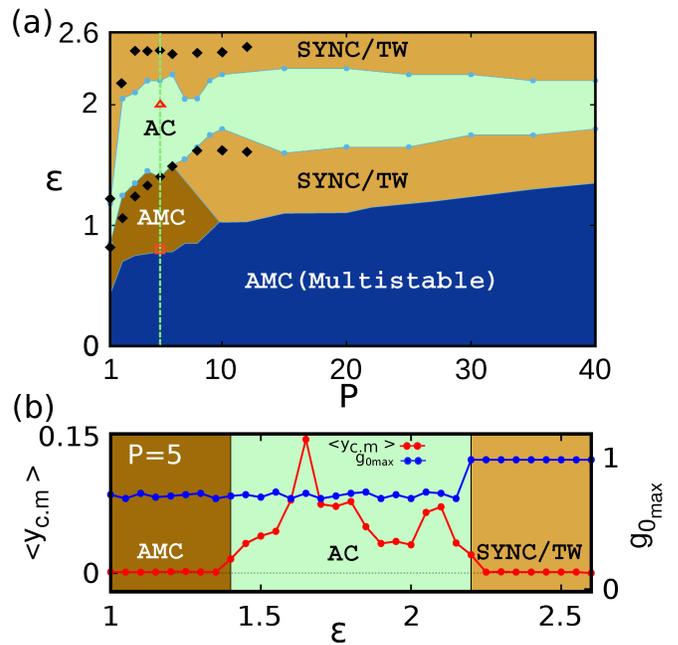}
\caption{\label{phase} (Color online) (a) Dynamic regimes in the ($P,\varepsilon$) parameter spacefor $N=200$. AMC: Amplitude-mediated phase chimera, AC: Amplitude chimera, SYNC \& TW: Synchronized and/or coherent traveling wave solution, AMC (Multistable): AMC state coexists with SYNC \& TW. $\Box$ and $\Delta$ denote the parameter values used in Fig.~\ref{fig_amc} (for the AMC state) and Fig.~\ref{fig_ac} (for the AC state), respectively. $\blacklozenge$ indicates the pitchfork bifurcation points (PB1 and PB2) computed using XPPAUT (see Sec.~\ref{symbreak} for a detailed discussion). The light blue dots on the edges of the AC region indicate the threshold values of $\varepsilon$ and $P$ with $\left<y_{c.m}\right>>0$ and $g_{0max}<1$. (b) Mean center of mass coordinate $\left<y_{c.m}\right>$ and spatial correlation measure $g_{0max}$ for $N=200$ and $P=5$, i.e., along the vertical dashed line of (a). See Table I and text for details. Other parameters are $\omega=2$, $\delta=1$, $\phi=\pi/2-0.1$.}
\end{figure}

All the above prominent spatiotemporal patterns are mapped in the diagram of dynamic regimes in Fig.\ref{phase} (a) in the ($P,\varepsilon$) plane. 
To identify different zones in the phase diagram we follow the criteria shown in Table I. Here $\left<y_{c.m}\right>$ is defined as
\begin{equation}
\left<y_{c.m}\right>=\frac{\sum_{i=1}^{N}|y_{c.m_i}|}{\mbox{max}\{n,1\}}.
\end{equation} 
where $y_{c.m_i}$ is given by Eq.~(\ref{cm}), $N$ is the total number of nodes, $n$ is the number of nodes with shifted center of mass of the oscillation (i.e., the number of nodes in the incoherent region). Note that for an AC state $\left<y_{c.m}\right>>0$ whereas $\left<y_{c.m}\right>=0$ for the AMC and synchronized or coherent traveling wave (SYNC/TW) states. As for a chimera state $g_0<1$ and for a globally synchronized state $g_0=1$, we distinguish chimera and SYNC/TW states by using $g_0$. However, since for chimera states $g_0$ shows fluctuation around an average value, to avoid any ambiguity we use the maximum value of $g_0$ denoted as $g_{0max}$.
\begin{table}
\caption{Criteria for identifying different dynamic regimes in Fig.\ref{phase}}
\label{table}
\begin{tabular}{|c|c|}
\hline
{\bf Observations} & {\bf Condition}\\
\hline
\hline
 AC & $\left<y_{c.m}\right>>0$ AND $g_{0max}<1$\\
\hline
AMC & $\left<y_{c.m}\right>=0$ AND $g_{0max}<1$\\
\hline
SYNC/TW & $\left<y_{c.m}\right>=0$ AND $g_{0max}=1$\\
\hline
\end{tabular}
\end{table}
From the phase diagram it is observed that for a given coupling range ($P$) AMC occurs at a lower coupling strength $\varepsilon$ and with increasing $\varepsilon$, beyond a certain value of $\varepsilon$, AC emerges. Significantly, for a lower value of $P$ we observe a direct transition from AMC to AC with increasing $\varepsilon$. 
This may be due to the fact that typically in a network, larger $P$ favors multistability, therefore, an increase in $P$ may promote the region of multistability by suppressing the pure AMC region, making it difficult to observe direct transitions from AMC to AC.\footnote{We have verified that the direct transition from AMC to AC occurs up to a maximum coupling range  $P_{max}$ such that $0.025 \le \frac{P_{max}}{N} \le 0.05$.}
The phase diagram of Fig.~\ref{phase} (a) demonstrates that the direct transition occurs for $P\le5$. This direct transition from AMC to AC and then to SYNC/TW is illustrated clearly in Fig.~\ref{phase} (b) for $P=5$. With increasing $\varepsilon$, for $\varepsilon<1.35$, $\left<y_{c.m}\right>=0$ indicating that all the oscillators are oscillating around the origin, however, $g_{0max}<1$ indicates that it is a chimera state: therefore in this region the system shows an AMC state. In the range $1.35<\varepsilon<2.25$ the system has $\left<y_{c.m}\right>>0$ indicating the presence of shifted center of mass limit cycles in the spatiotemporal pattern and additionally $g_{0max}<1$ confirms that in this parameter regime the system indeed shows amplitude chimeras. Finally, for $\varepsilon>2.25$ the network shows $\left<y_{c.m}\right>=0$ (indicating that all the nodes are oscillating around the origin) and $g_{0max}=1$ (indicating the absence of the coexistence of synchrony and asynchrony), therefore, this region belongs to the SYNC/TW state. We also check for the presence of hysteresis during this transition from one chimera state to another but could not detect any. For larger coupling range ($P$) a traveling wave or synchronized pattern (SYNC) is interspersed between AMC and AC. We observe that for small $\varepsilon$ and large $P$ the AMC state is multistable and coexists with the fully synchronized oscillations (SYNC) or coherent traveling waves (TW), this is shown in Fig.~\ref{phase} (a).

\begin{figure}
\centering
\includegraphics[width=0.48\textwidth]{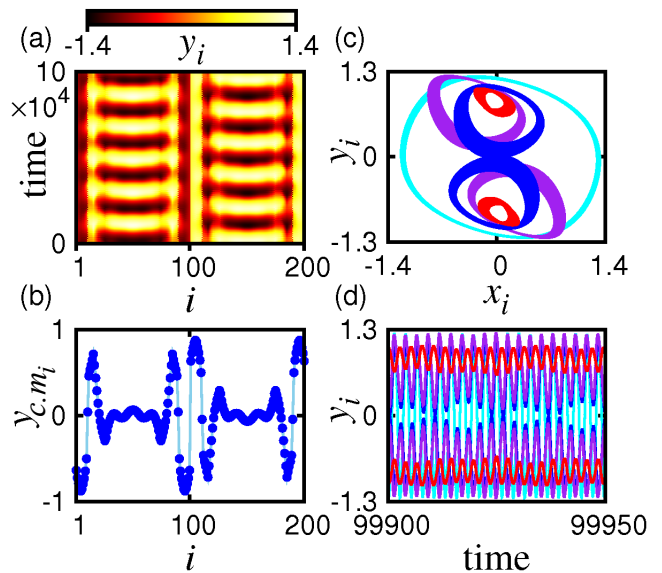}
\caption{\label{fig_vac} (Color online) Variable-amplitude chimera (VAC) for $P=15$ and $\varepsilon=1.67$. (a) Space-time diagram of $y_i$. (b) Corresponding center of mass ($y_{c.m_i}$). (c) Phase portrait in the ($x_i-y_i$) plane of a few selected oscillators. (d) Corresponding time series of $y_i$. Other parameters: $\omega=2$, $\delta=1$, $\phi=\pi/2-0.1$.}
\end{figure}

Moreover, several other chimera patterns are observed in narrow regions of the parameter space (not shown in Fig.\ref{phase} (a)); the most prominent one is the variable amplitude AC. Typically, in an AC state all the oscillators exhibit {\it periodic} limit cycle oscillations with the same phase velocity, however, in our case, in a parameter region near the transition from AMC to AC and from AC to SYNC we observe quasiperiodic oscillations in the AC state. We name this state as the {\it variable amplitude chimera} (VAC). Figure~\ref{fig_vac} (a) demonstrates the spatiotemporal pattern of the VAC for $P=15$ and $\varepsilon=1.67$, and Fig.~\ref{fig_vac} (b) shows the corresponding centers of mass ($y_{{c.m}_i}$) of each oscillator. One can also visualize the apparently quasiperiodic variation in amplitude from the phase portrait and the corresponding time series in Fig.~\ref{fig_vac}~(c) and Fig.~\ref{fig_vac}~(d), respectively. 

\subsection{Qualitative explanation: Symmetry-breaking bifurcations}\label{symbreak}
Next, we try to understand the observed phenomena qualitatively in the light of bifurcation scenarios. However, since we are considering a large network of coupled oscillators with amplitude dynamics, it is difficult to reveal the complete bifurcation structure in such a high dimensional phase space, and connect it to our observations of chimera patterns. Therefore, we start from a smaller network and then systematically attempt to find the connection between the observed chimera patterns and the relevant bifurcation mechanism of the complete network. 

\begin{figure}
\centering
\includegraphics[width=0.48\textwidth]{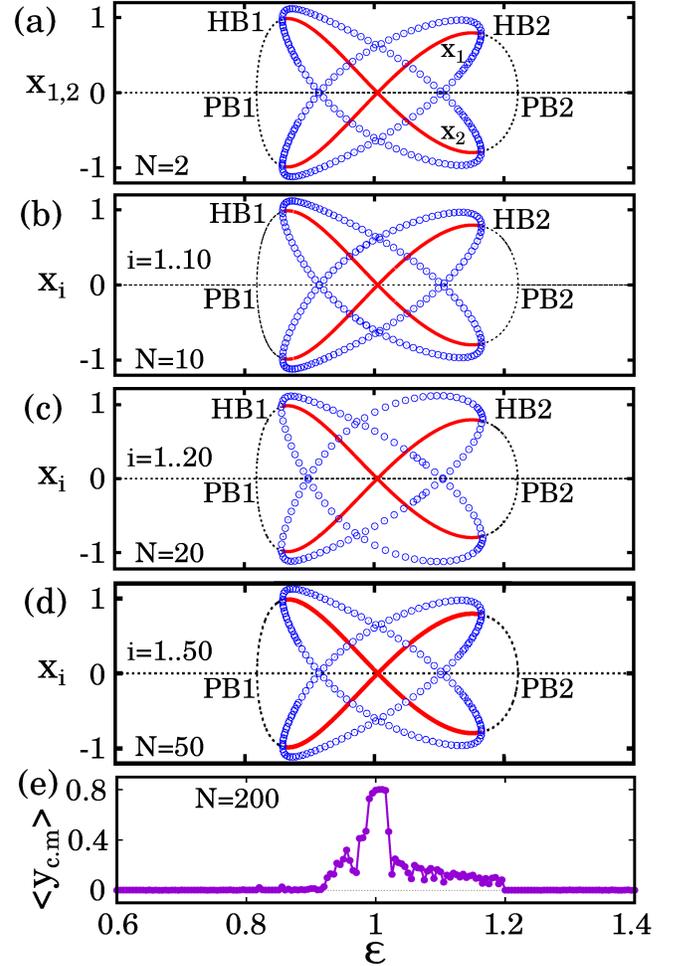}
\caption{\label{bif} (Color online) Bifurcation diagram (using XPPAUT) of coupled Rayleigh oscillators under matrix coupling (Eq.~\ref{ral}) with coupling range $P=1$ for (a) $N=2$, (b) $N=10$, (c) $N=20$, and (d) $N=50$. Periodic orbits emanating from subcritical Hopf bifurcations are shown in open (blue) circles for only the first oscillator $i=1$, however, the fixed point solution of {\it all} the oscillators are shown and they are lying on top of each other. (e) $\left<y_{c.m.}\right>$ of $N=200$: Non-zero value indicates the appearance of the AC state. Red thick lines: stable fixed points; dashed black lines: unstable fixed points; open circles (light blue): unstable limit cycles. PB1, and PB2: Pitchfork bifurcation points; HB1 and HB2: Subcritical Hopf bifurcation points. Parameters are: $\phi=\frac{\pi}{2}-0.1$, $\delta=1$ and $\omega=2$.}
\end{figure}

We start by considering two Rayleigh oscillators coupled via matrix coupling, and derive the bifurcation points. Equation.~\eqref{ral} is rewritten for two oscillators:
\begin{subequations}
\label{rayos}
\begin{align}
\dot{x}_{1,2}&=\omega y_{1,2}\nonumber\\
&+\varepsilon \big(a_{11}(x_{2,1}-x_{1,2})+a_{12}(y_{2,1}-y_{1,2})\big)\\
\dot{y}_{1,2}&=-\omega x_{1,2}+\delta(1-y_{1,2}^2)y_{1,2}\nonumber\\
&+\varepsilon \big(a_{21}(x_{2,1}-x_{1,2})+a_{22}(y_{2,1}-y_{1,2})\big)
\end{align}
\end{subequations}
We will derive the results for the general matrix $A=\left( \begin{array}{cc} a_{11} & a_{12} \\ a_{21} & a_{22} \end{array} \right)$. The system of equations \eqref{rayos} has three fixed points: one trivial fixed point $(0,0,0,0)$ and a pair of nontrivial fixed points $(x^*,y^*,-x^*,-y^*)$ with 
\begin{subequations}
\label{ray_od}
\begin{align}
x^*&=\pm \sqrt{\frac{(\omega-2\varepsilon a_{12})^2}{4a_{11}^2\varepsilon^2}}y^*,\\
y^*&=\pm\sqrt{\frac{2\varepsilon a_{11}(\delta-2\varepsilon a_{22})+(2\varepsilon a_{12}-\omega)(2\varepsilon a_{21}+\omega)}{2\varepsilon\delta a_{11}}}.
\end{align}
\end{subequations} 
The linear stability analysis of the fixed points yields that with increasing $\varepsilon$ the unstable trivial fixed point undergoes a symmetry-breaking pitchfork bifurcation giving rise to two additional nontrivial unstable fixed points $(x^*,y^*)$ at $\varepsilon_{PB1}$,  
\begin{equation}
\label{epsi}
\varepsilon_{PB1}=\frac{\alpha-\sqrt{\beta}}{\Delta},	
\end{equation}
where $\alpha=-\delta a_{11}-(a_{12}-a_{21})\omega$, $\beta=\Delta\omega^2+(\delta a_{11}+(a_{12}-a_{21})\omega)^2$ and $\Delta=4(a_{12}a_{21}-a_{11}a_{22})$. Three fixed points (one trivial and two nontrivial ones) collide at $\varepsilon_{PB2}$ and symmetry reappears, where
\begin{equation}
\label{epsii}
\varepsilon_{PB2}=\frac{\alpha+\sqrt{\beta}}{\Delta},	
\end{equation}
Therefore, between $\varepsilon_{PB1}$ and $\varepsilon_{PB2}$ a bubble-like symmetry-breaking inhomogeneous steady states (i.e., oscillation death state) emerges~\cite{kosprl,ZAK13a,tanpre1,SCH15b}. This scenario is shown in Fig.~\ref{bif}(a) for two oscillators (here we show the $x$ variable, however, $y$ variable also gives the similar qualitative bifurcation structure). Using $a_{11}=a_{22}=\cos\phi$ and $a_{12}=-a_{21}=\sin\phi$. For $\phi=\frac{\pi}{2}-0.1$, $\delta=1$ and $\omega=2$ we get $\varepsilon_{PB1}=0.818$ and $\varepsilon_{PB2}=1.221$. Next, we search for the Hopf bifurcation points, which can be computed from the two dominant eigenvalues of the Jacobian of the nontrivial fixed points of \eqref{rayos},
\begin{equation}
\lambda_{1,2}=\frac{-\mu\pm\sqrt{\mu^2-(4\varepsilon\omega a_{11})^2}}{4\varepsilon\omega a_{11}},
\end{equation}
where $\mu=4\varepsilon a_{11}(\delta-3\varepsilon a_{22})+3(2\varepsilon a_{12}-\omega)(2\varepsilon a_{21}+\omega)$. From this expression using the above parameter values we have $\varepsilon_{HB1}=0.858$ and $\varepsilon_{HB2}=1.165$, which agrees well with the numerical bifurcation diagram of Fig.~\ref{bif}(a). Therefore, with increasing $\varepsilon$, beyond  $\varepsilon_{PB1}$, the unstable inhomogeneous fixed point branches are stabilized through a subcritical Hopf bifurcation at $\varepsilon_{HB1}$ and again become unstable at $\varepsilon_{HB2}$ through an inverse subcritical Hopf bifurcation. Between $\varepsilon_{HB1}$ and $\varepsilon_{HB2}$ these fixed points are accompanied by unstable limit cycles with shifted center of mass of the oscillations (that are the characteristics of amplitude chimeras), and also by synchronous oscillations.

\begin{figure}
\centering
\includegraphics[width=0.48\textwidth]{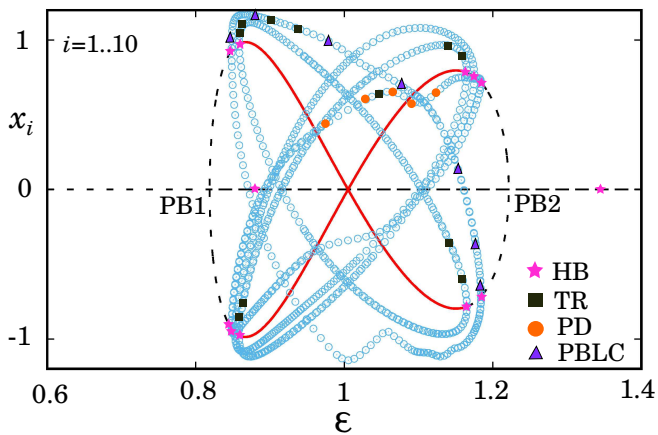}
\caption{\label{bif10} (Color online) Bifurcation diagram (using XPPAUT) of $N=10$ coupled Rayleigh oscillators under matrix coupling (Eq.~\ref{ral}) with $\phi=\frac{\pi}{2}-0.1$. Red thick lines: stable fixed points; dashed black lines: unstable fixed points; open circles (light blue): unstable limit cycles. PB1, and PB2: Pitchfork bifurcation points; HB: Hopf bifurcation, TR: Torus bifurcation, PD: Period-doubling bifurcation, PBLC: Pitchfork bifurcation of limit cycles. Periodic orbits emanating from ten Hopf bifurcation points are shown in open circles for only the first oscillator $i=1$, however, the fixed point solutions of {\it all} the oscillators are shown and they are lying on each other. Parameters are: $\delta=1$, $\omega=2$.}
\end{figure}

Now we consider $N>2$ and interestingly find that for any $N$ the pitchfork bifurcation points PB1 and PB2 are the same as given by Eq.~(\ref{epsi}) and Eq.~(\ref{epsii}), respectively, as long as we consider nearest neighbor coupling (i.e., $P=1$). This is due to the fact that an oscillator `sees' the same environment for a nearest neighbor coupling. Figs.~\ref{bif}(b), (c) and (d) show this for $N=10, 20,$ and $50$, respectively. However, as $N$ increases, a large number of additional Hopf points appear between PB1 (HB2) and HB1 (PB2), and each Hopf point gives rise to additional unstable limit cycles around the nontrivial fixed points. In Fig.~\ref{bif}(b) ($N=10$), Fig.~\ref{bif}(c) ($N=20$), and Fig.~\ref{bif}(d) ($N=50$) we only show the unstable limit cycles created through Hopf bifurcations at HB1 and HB2 on the upper and lower branches (for clarity only the orbits of a single oscillator with $i=1$ is shown). It is interesting to note that the limit cycle created on the upper (lower) branch at $\varepsilon_{HB1}$ (left side) terminates on the lower (upper) branch at $\varepsilon_{HB2}$ (right side). Therefore, in this system we have a localized region between PB1 and PB2 where a large number of unstable limit cycles with shifted center of mass are ``trapped" and therefore coexist in a broad region (or hypervolume) of the phase space. Also, note that in this parameter region (stable) limit cycles around the trivial fixed point (which is the origin) still coexist with the shifted limit cycles. This coexistence of limit cycles with shifted center of mass and in-phase oscillations without shifted center of mass may be attributed to the existence of amplitude chimeras. To demonstrate the complexity of the dynamical behavior in the ``trapped" region we show some representative orbits and bifurcation points in Fig.~\ref{bif10} for $N=10$. Out of twenty-two Hopf bifurcation points which we have identified (using XPPAUT), here we show only the unstable orbits of the oscillator with $i=1$ emanating from ten Hopf bifurcation points (shown with open circles). Additionally, the (secondary) bifurcation of limit cycles makes the scenario much more complex; we identify torus bifurcations (TR), period doubling bifurcations (PD) and pitchfork bifurcations of limit cycle (PBLC) (see Fig.~\ref{bif10}). The presence of torus bifurcations (TR) and period doubling bifurcations (PD)  may be responsible for the variable-amplitude AC (VAC) state where the  limit cycles with shifted center of mass are either quasiperiodic (see Fig.~\ref{fig_vac}) or higher periodic in nature. 

The next question arises: is our argument that ACs always appear in the symmetry broken ``trapped" region between PB1 and PB2, also true for larger network size? We find that AC indeed appears in the ``trapped" region even for larger networks. This is shown in Fig.~\ref{bif} (e) for $N=200$ and $P=1$: $\left<y_{c.m.}\right>>0$ indicates an AC state, which appears between $\varepsilon_{PB1}$ and $\varepsilon_{PB2}$ of the smaller networks with nearest neighbor coupling (see Fig.~\ref{bif}(a)--(d)). We have checked this also for much larger network sizes with $N=500$ and $N=1000$ and have obtained the same result.  

In the above discussion we have considered nearest neighbor coupling (i.e., $P=1$). Next, we extend our bifurcation analysis to an arbitrary coupling range $P$. For this we consider the network with $N=200$ and compute the bifurcation points (using XPPAUT) for different coupling ranges ($P$). In this case, too, we locate two pitchfork bifurcation points PB1 (where symmetry is broken) and PB2 (where symmetry is restored). These points are shown in the phase diagram of Fig.~\ref{phase} (a) using $\blacklozenge$ symbols: in the phase diagram, the PB1 points are below the AC region and the PB2 points are above the AC region. We plot the results only up to $P=12$ because for $P>10$, $\varepsilon_{PB1}$ and $\varepsilon_{PB2}$ do not change appreciably with $P$. It is important to note that the AC region always lies in between PB1 and PB2 (i.e., the ``trapped" region) for any coupling range, confirming the connection of symmetry-breaking bifurcations with the emergence of AC. However, it is noteworthy that the AC region is narrower inside this trapped region (specially for $P>2$). This is due to the fact that the exact region of appearance of unstable periodic orbits is governed by the Hopf bifurcations on the symmetry-breaking fixed point branches, and this region is narrower than the range between PB1 and PB2. Due to the large size of the network, the continuation package fails to provide the exact location of Hopf points and the shape of the limit cycles emanating from those points inside this region. 

Finally, we try to understand the mechanism behind the long lifetime of the observed AC. In the earlier cases where AC was observed in Stuart-Landau oscillators with nonlocal coupling \cite{scholl_CD}, there exists only one symmetry breaking pitchfork bifurcation (PB) point beyond which symmetry breaks (see Appendix B). In that case the oscillations with shifted center of mass (i.e., the incoherent oscillation) are unstable limit cycle oscillations emerging from a subcritical Hopf bifurcation on the symmetry breaking fixed point branches and these center of mass-shifted oscillations always coexist with the in-phase oscillations. Therefore, if a certain node in the network starts as an center of mass-shifted oscillator, due to the unstable nature of the limit cycle, after a certain time it eventually ends up with the in-phase synchronized members of the network: this makes AC in Stuart-Landau oscillators with nonlocal coupling \cite{scholl_CD} a relatively short-living chimera pattern. The detailed study of the lifetime of AC states in Stuart-Landau oscillators is reported in Refs.~\onlinecite{ac_noise,TUM17}. Note that in case of Stuart-Landau oscillators with symmetry-breaking nonlocal coupling large lifetimes can arise for certain values of the coupling range and strength due to the phase space structure, and they have been explained by a Floquet stability analysis \cite{TUM17}. In the present case, although the center of mass-shifted limit cycles are unstable, however, they are always {\it trapped} in between two symmetry-breaking bifurcation points PB1 and PB2. As a result, the system has a large number of dense unstable limit cycles concentrated in a localized region of phase space. Therefore, if a node starts on (or near) an unstable orbit (depending upon initial conditions), there always exist nearby unstable orbits that act like a saddle to force the node to stay near that trajectory. This makes the lifetime of the center of mass-shifted limit cycle (and hence the AC) appreciably long. Intuitively, the number of unstable limit cycles in the ``trapped" region increases with increasing network size, therefore, we should obtain an increasing lifetime with increasing $N$. In fact, we find that even with $N=20$ the resulting AC has a very long lifetime: we checked it for a simulation time of $10^7$ and still observed a stable AC pattern (Appendix B). A long-living amplitude chimera in a small network is itself an important observation and it supports our argument of connection between long-living AC and the presence of localized dense unstable periodic orbit in a ``trapped" parameter region.      

Therefore, based on the above observations we make the following two conjectures: (i) The existence of symmetry breaking bifurcations of the fixed points and the presence of Hopf bifurcations on the symmetry-breaking fixed-point branches are {\it necessary} (if not sufficient) to observe an AC state, (ii) The existence of a large number of close dense unstable periodic orbits in a trapped (or localized) region of parameter space (and phase space) is crucial for the long lifetime of an AC state.

\section{Conclusion}
We have reported the observation of both amplitude-mediated phase chimeras and amplitude chimeras in a single network of coupled identical oscillators. This provides a bridge between two distinct chimera states. We have shown that for small coupling range a direct transition from AMC state to AC state occurs. We have further given evidence that the amplitude chimera is not a short-living transient spatiotemporal pattern, rather it has a long lifetime. Recently, Gjurchinovski {\em et al.}~\cite{delay_ac} have used time-delay to stabilize the amplitude chimera state in a network of Stuart-Landau oscillators, but here we do not use any control scheme, rather the long-living amplitude chimera state appears naturally. Also, apart from periodic temporal oscillations we have also found quasiperiodic (or higher periodic) oscillations in the incoherent part of the amplitude chimera. 

We have also raised the issue, why some oscillators show amplitude-mediated phase chimeras and others exhibit amplitude chimeras. Our study indicates that amplitude chimeras occur only above a certain critical coupling strength where symmetry-breaking pitchfork  bifurcations of nontrivial inhomogeneous steady states take place. 
We further intuitively identify the role of closely separated dense unstable orbits trapped in a region of phase space in governing the lifetime of amplitude chimeras. This region interspersed between two symmetry breaking bifurcations in parameter space arises due to the interplay of the local dynamics of the Rayleigh oscillator and the particular form of the coupling matrix. We did not observe this type of trapped region in the case of Rayleigh oscillators with nonlocal diffusive coupling. Therefore, in those cases the amplitude chimeras are found to be short-living spatiotemporal patterns.     

Since the two chimera states emerge due to different types of symmetry-breaking phenomena \cite{kemethprl}, therefore  our finding of a continuous transition from AMC to AC will be important to understand the connection between the two variants of symmetry-breaking state. Also, in robotics, Rayleigh oscillators are used to model human limb movement and locomotion; see for example Ref.~\onlinecite{robot}, which discusses how a bipedal robot can be modeled by using mutually coupled Rayleigh oscillators. Therefore, apart from improving the fundamental understanding of the chimera state, our results may be relevant for robotics \cite{robot}.

\begin{acknowledgments}
E.S. and A.Z. acknowledge the financial support by DFG in the framework of SFB 910.
\end{acknowledgments}

\appendix
\section{Completely random initial conditions}
\begin{figure}
\centering
\includegraphics[width=0.42\textwidth]{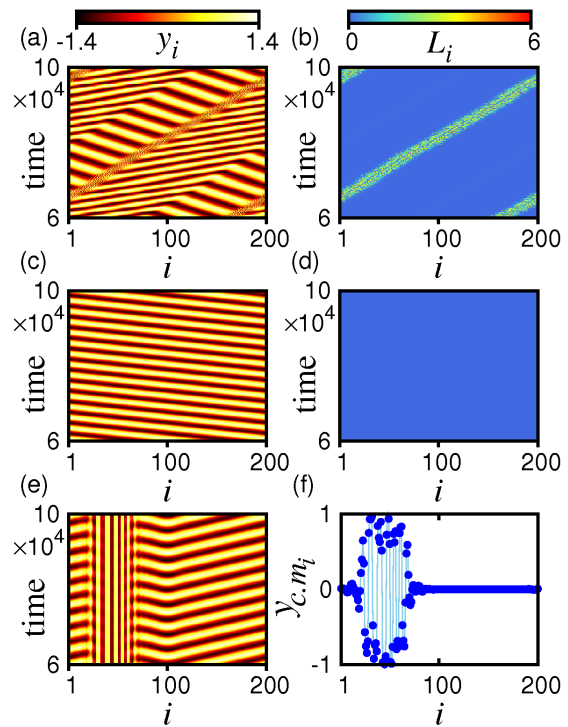}
\caption{\label{ranac} (Color online) Completely random initial condition, for coupling range $P=5$: (a) Amplitude-mediated phase chimera (traveling) and (b) its local curvature ($L_i$) for $\varepsilon=0.85$. (c) Coherent traveling wave and (d) corresponding $L_i$ for $\varepsilon=1.3$. (e) Amplitude chimera and (f) its center of mass ($y_{c.m_i}$) for $\varepsilon=1.58$. Parameters are: $\delta=1$, $\omega=2$, $\phi=\pi/2-0.1$.}
\end{figure}
Here we verify our results with completely random initial conditions and find qualitatively similar scenarios as discussed in the main text. For an exemplary illustration we choose $P=5$ (as in Fig.~\ref{fig_amc} and Fig.~\ref{fig_ac}) and consider random initial condition uniformly distributed in $x,y\in(-0.5,0.5)$. We observe that with increasing coupling strength $\varepsilon$ the network undergoes a transition from AMC to traveling wave and finally to AC. Figure~\ref{ranac} shows the transition scenario AMC ($\varepsilon=0.85$) [Fig.~\ref{ranac}(a)] to TW ($\varepsilon=1.3$) [Fig.~\ref{ranac}(c)] and finally to AC ($\varepsilon=1.58$) [Fig.~\ref{ranac}(e)].  Figure~\ref{ranac}~(b) and Fig.~\ref{ranac}~(d) depict the plots of the local curvature $L_i$ indicating the occurrence of AMC and TW, respectively. Also, the plot of the center of mass ($y_{c.m_i}$) 
of each oscillator corresponding to Fig.~\ref{ranac}~(e) is shown in Fig.~\ref{ranac}~(f) ensuring the occurrence of AC. Note that the AMC state here is actually a traveling AMC and also we do not find any direct transition from AMC to AC, but  rather an intermediate TW state. Nevertheless, the occurrence of AMC and AC for completely random initial conditions indicates the generality of the  phenomenon. 

\section{Diffusive coupling: A single symmetry-breaking bifurcation point}
In the main text we have shown that the matrix coupling in a network of Rayleigh oscillators gives rise to multiple symmetry-breaking bifurcations. In contrast, here we will show that a diffusive coupling in Rayleigh oscillators as well as Stuart-Landau oscillators gives rise to a single symmetry-breaking bifurcation. Two Rayleigh oscillators coupled through diffusive coupling via the $x$ variable is given by
\begin{subequations}
\label{ros1}
\begin{align}
\dot{x}_{1,2} &=\omega y_{1,2}+\varepsilon(x_{2,1}-x_{1,2})\\
\dot{y}_{1,2} &=-\omega x_{1,2}+\delta (1-{y_{1,2}}^2)y_{1,2}.
\end{align}
\end{subequations}
The trivial unstable fixed point is $(0,0,0,0)$. A pair of nontrivial unstable fixed points $(x^*,y^*,-x^*,-y^*)$  with
$x^*=\pm \frac{\omega}{2\varepsilon}\sqrt{1-\frac{{\omega}^2}{2\varepsilon \delta}}$ and $y^*=\frac{2\varepsilon x^*}{\omega}$ 
appears through a pitchfork bifurcation for $\varepsilon > \varepsilon_{PB}$: ${\varepsilon}_{PB} = \frac{{\omega}^2}{2\delta}$. The unstable inhomogeneous fixed points $(x^*,y^*,-x^*,-y^*)$ are stabilized in a subcritical Hopf bifurcation at ${\varepsilon}_{HBS} = \frac{3{\omega}^2}{4\delta}$.
\begin{figure}
\centering
\includegraphics[width=0.49\textwidth]{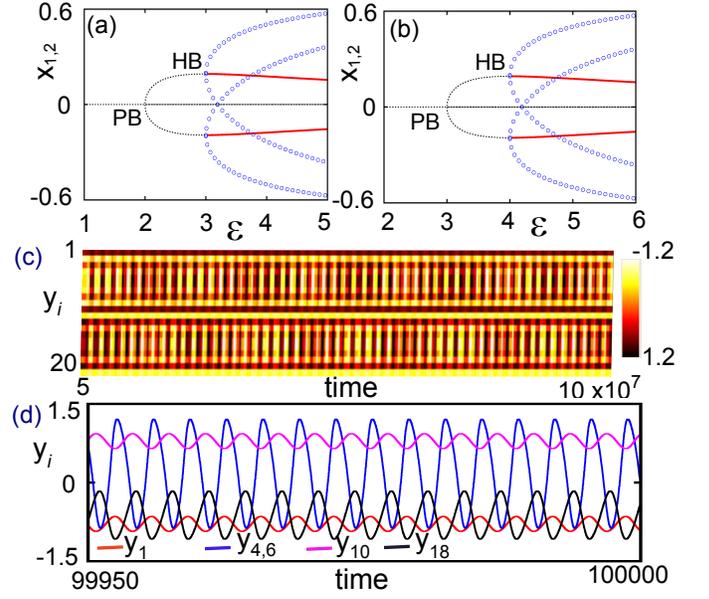}
\caption{\label{symbif} (Color online) Bifurcation diagram of two diffusively coupled (a) Rayleigh oscillators (Eq.~\ref{ros1}) and (b) Stuart-Landau oscillators (Eq.~\ref{sl1}); PB: pitchfork bifurcation, HB: Hopf bifurcation; Red thick lines: stable fixed points; dashed black lines: unstable fixed points; open circles (blue): unstable limit cycles. (c),(d): Amplitude chimeras of $N=20$ Rayleigh oscillators under matrix coupling of Eq.~(\ref{ral}) for $P=1$ and $\varepsilon=1$, (c) spatiotemporal plot, (d) time series of a few incoherent ($y_{1,10,18}$) and coherent ($y_{4,6}$) nodes. Other parameters are: $\delta=1$, $\omega=2$, $\phi=\pi/2-0.1$.}
\end{figure}
For Stuart-Landau oscillators under diffusive coupling the equation reads 
\begin{subequations}
\label{sl1}
\begin{align}
\dot{x}_{1,2} &=(1-{x_i}^2-{y_i}^2)x_{1,2}-\omega y_{1,2}+\varepsilon(x_{2,1}-x_{1,2})\\
\dot{y}_{1,2} &=\omega x_{1,2}+(1-{x_i}^2-{y_i}^2)y_{1,2}.
\end{align}
\end{subequations}
This equation is the limiting case (i.e, $N=2$ oscillator case) of the equation studied by Zakharova {\it et al.} \cite{scholl_CD} where the notion of the amplitude chimera was discovered. Also, Eq.~\eqref{sl1} was studied in detail by Koseska {\it et al.}\cite{kosprl} and Zakharova {\it et al.}\cite{ZAK13a} where they showed that a single symmetry-breaking bifurcation occurs at $\varepsilon=\frac{1+\omega^2}{2}$ and the symmetry-breaking fixed point branches are stabilized through a subcritical Hopf bifurcation. A detailed analytical and numerical study of 
large networks of nonlocally coupled Stuart-Landau oscillators with symmetry-breaking coupling was performed in Ref.~\onlinecite{SCH15b}, where a family of inhomogeneous steady states (oscillation death) and various multicluster patterns were found.

The bifurcation scenario of two diffusively coupled Rayleigh oscillators (Eq.~\ref{ros1}) is shown in Fig.~\ref{symbif} (a) and that of two Stuart-Landau oscillators (Eq.~\ref{sl1}) is shown in Fig.~\ref{symbif} (b). Both bifurcation diagrams show that after a pitchfork bifurcation (PB) unstable limit cycles arise from (subcritical) Hopf bifurcations. We also check our result for a larger number of oscillators with nonlocal diffusive coupling originally used in Ref.~\onlinecite{scholl_CD} and find that the number of pitchfork bifurcation points remains the same. In contrast to our case of Rayleigh oscillators with matrix coupling (Eq.~(\ref{ral})), in none of these cases further subcritical Hopf bifurcations generating further unstable limit cycles are detected: therefore, in these networks one does not have a region of dense localized unstable limit cycles, and in consistency with our argument we obtain relatively short-living amplitude chimeras. 

As mentioned in the main text, we obtain long-living amplitude chimeras with matrix-coupled Rayleigh oscillators (Eq.~\ref{ral}) even for small network sizes, e.g., $N=20$. This is shown in Fig.~\ref{symbif} (c) and (d) with $\varepsilon=1$ and $P=1$ (other parameters as in Fig.~\ref{fig_ac}). From the spatiotemporal plot of Fig.~\ref{symbif} (c) we observe a long-living AC (we limit our simulation time to $10^7$). Figure~\ref{symbif} (d) shows the representative time series of a few  incoherent nodes (i.e., oscillations with shifted center of mass, e.g., $y_{1,10,18}$) and coherent nodes (i.e., oscillations without shifted center of mass, e.g., $y_{4,6}$), which characterizes the AC state in the system.




%
\end{document}